\documentstyle[12pt]{article}

\newcommand{\E}{{\cal{E}}}

\renewcommand{\d}{{\rm d}}
\newcommand{\I}{{\rm i}}
\renewcommand{\a}{\alpha}
\newcommand{\dfrac}[2]{\displaystyle\frac{#1}{#2}}
\newcommand{\be}{\begin{equation}}
\newcommand{\ee}{\end{equation}}
\newcommand{\bea}{\begin{eqnarray}}
\newcommand{\eea}{\end{eqnarray}}
\def\J#1#2#3#4{(#1). {#2} {\bf #3}, #4}
\def\GRG{\it Gen. Relat. Grav.}

\def\PTP{\it Prog. Theor. Phys.}

\def\PRD{{\it Phys. Rev.} D}
\def\PR{\it Phys. Rev.}

\def\APNY{\it Ann. Phys. (N.Y.)}

\def\JMP{\it J. Math. Phys.}
\def\CQG{\it Class. Quantum Grav.}

\def\PLA{\it Phys. Lett. A}
\def\BAPS{\it Bull. Acad. Polon. Sci. Ser. Math. Astron. Phys.}
\def\BSMB{\it Bull. Soc. Math. Belg.}

\begin{document}
\title{Demia\'nski--Newman Solution Revisited}
\author{J.~A.~Aguilar--S\'anchez,\dag\, A.~A.~Garc\'\i a\ddag\,
and V.~S.~Manko\ddag}
\date{}
\maketitle

\vspace{-1cm}

\begin{center}
\dag Departamento de F\'\i sica, UAEMex, C.P. 50000, Toluca,
Mexico\\
\medskip
\ddag Departamento de F\'\i sica, Centro de Investigaci\'on y de
Estudios\\ Avanzados del IPN, A.P. 14-740, 07000 M\'exico D.F.,
Mexico
\end{center}

\vspace{.1cm}

\begin{abstract}
The derivation of the Demia\'nski--Newman solution within the
framework of the Ernst complex formalism is considered. We show
that this solution naturally arises as a two--soliton
specialization of the axisymmetric multi--soliton electrovacuum
metric, and we work out the full set of the corresponding metrical
fields and electromagnetic potentials. Some limits and physical
characteristics of the DN space--time are briefly discussed.
\end{abstract}

\section{Introduction}

The Demia\'nski--Newman (DN) metric \cite{DN} is a five--parameter
stationary axisymmetric solution of the Einstein--Maxwell
equations which generalizes the well--known Kerr--Newman spacetime
\cite{KN}, its two additional parameters being the gravitomagnetic
and magnetic monopoles; it is in turn a special case of some more
general metrics \cite{PD} widely discussed in the literature. The
DN metric was originally derived with the aid of the complex
coordinate transformation procedure, so it seems likely to have
its representation within the framework of the Ernst formalism
\cite{E} because this solution turns out to be the most general
two--soliton specialization of the electrovacuum multi--soliton
metric \cite{RMM}, constituting a basic element of the stationary
systems of aligned charged, magnetized, spinning particles. In
Section~2 we shall obtain the `canonical' form of the Ernst
complex potentials for the DN solution involving the analytically
extended parameter set, the corresponding metric functions and
electromagnetic potentials. In Section~3 we shall discuss some
properties of the DN spacetime. Section~4 contains concluding
remarks.

\section{The Ernst potentials of the DN solution, the corresponding
metric functions and electromagnetic potentials}

The Ernst complex potentials $\E$ and $\Phi$ \cite{E} defining the
DN solution arise from the axis data of the form \be
\E(\rho=0,z)=\frac{z-m-\I(a+\nu)}{z+m-\I(a-\nu)}, \quad
\Phi(\rho=0,z)=\frac{q+\I b}{z+m-\I(a-\nu)}, \ee where $\rho$ and
$z$ are the Weyl--Papapetrou cylindrical coordinates, and the five
arbitrary parameters $m$, $a$, $\nu$, $q$, $b$ are the total mass,
total angular momentum per unit mass, gravitomagnetic monopole
(NUT parameter \cite{NTU}), electric and magnetic charges,
respectively. This interpretation of the parameters follows from
the consideration of the Simon multipole moments \cite{Sim}
corresponding to the axis data (1) with the aid of the
Hoenselaers--Perj\'es procedure \cite{HP} which gives the
following expressions for the first four moments ($M_i$, $J_i$,
$Q_i$, $B_i$ stand, respectively, for the mass, angular momentum,
electric and magnetic moments): \bea &&M_0=m, \quad M_1=-\nu a,
\quad M_2=-ma^2, \quad M_3=\nu a^3, \nonumber\\ &&J_0=\nu, \quad
J_1=ma, \quad J_2=-\nu a^2, \quad J_3=-ma^3, \nonumber\\ &&Q_0=q,
\quad Q_1=-ba, \quad Q_2=-qa^2, \quad Q_3=ba^3, \nonumber\\
&&B_0=b, \quad B_1=qa, \quad B_2=-ba^2, \quad B_3=-qa^3. \eea

The axis data (1) is the $N=1$ specialization of the axis
expressions of the electrovacuum axisymmetric solution \cite{RMM}
obtained with the aid of Sibgatullin's method \cite{S}, so we can
use the results of the paper \cite{RMM} for writing out the
potentials $\E(\rho,z)$, $\Phi(\rho,z)$ and all the metric fields
of the DN solution. Then for the potentials $\E$ and $\Phi$ we
have
\[ \E=E_+/E_-, \quad \Phi=F/E_-, \]
\be E_\pm=\left|
\begin{array}{ccc}
1 & 1 & 1 \vspace{0.15cm} \\ \pm 1 & \dfrac{r_1} {\a_{1}-\beta} &
\dfrac{r_2}{\a_{2}-\beta} \vspace{0.15cm}\\ 0 &
\dfrac{h(\a_{1})}{\a_{1}-\bar{\beta}} &
\dfrac{h(\a_{2})}{\a_{2}-\bar{\beta}}
\end{array} \right|, \quad F=\left|
\begin{array}{cc}
f(\a_1) & f(\a_2)  \vspace{0.15cm} \\
\dfrac{h(\a_{1})}{\a_{1}-\bar{\beta}} &
\dfrac{h(\a_{2})}{\a_{2}-\bar{\beta}}
\end{array} \right|,
\ee where $r_n\equiv\sqrt{\rho^2+(z-\a_n)^2}$ (a bar over a symbol
means complex conjugation), and for the definitions of $\a_n$,
$\beta$, $h(\a_n)$ and $f(\a_n)$ we refer to \cite{RMM}. The
corresponding metric functions $f$, $\gamma$ and $\omega$ entering
the axisymmetric `canonical' line element
\be
\d s^2=f^{-1}\left[{\rm e}^{2\gamma}(\d\rho^2+\d
z^2)+\rho^2\d\varphi^2 \right]-f(\d t-\omega\d\varphi)^2, \ee are
given by the expressions
\[ f=\frac{E_+\bar E_-+\bar
E_+E_-+2F\bar F}{2E_-\bar E_-}, \quad {\rm
e}^{2\gamma}=\frac{E_+\bar E_-+\bar E_+E_-+2F\bar F} {2K_0\bar
K_0r_1r_2},\]\[ \omega=\frac{2\,{\rm Im}\{ E_-\bar H-\bar
E_-G-F\bar I\}} {E_+\bar E_-+\bar E_+E_-+2F\bar F},\]
\[ G=\left|
\begin{array}{cc}
r_1+\a_1-z & r_2+\a_2-z  \vspace{0.15cm} \\
\dfrac{h(\a_{1})}{\a_{1}-\bar{\beta}} &
\dfrac{h(\a_{2})}{\a_{2}-\bar{\beta}}
\end{array} \right|, \quad H=\left|
\begin{array}{ccc}
z & 1 & 1 \vspace{0.15cm} \\ -\beta & \dfrac{r_1} {\a_{1}-\beta} &
\dfrac{r_2}{\a_{2}-\beta} \vspace{0.15cm}\\ \bar e &
\dfrac{h(\a_{1})}{\a_{1}-\bar{\beta}} &
\dfrac{h(\a_{2})}{\a_{2}-\bar{\beta}}
\end{array} \right|, \]
\be
I=\left|
\begin{array}{cccc}
f_1 & 0 & f(\a_1) & f(\a_2)  \vspace{0.15cm} \\ z & 1 & 1 & 1
\vspace{0.15cm} \\ -\beta & -1 & \dfrac{r_1} {\a_{1}-\beta} &
\dfrac{r_2}{\a_{2}-\beta} \vspace{0.15cm}\\ \bar e & 0 &
\dfrac{h(\a_{1})}{\a_{1}-\bar{\beta}} &
\dfrac{h(\a_{2})}{\a_{2}-\bar{\beta}}
\end{array} \right|, \quad K_0=\left|
\begin{array}{cc}
\dfrac{1} {\a_{1}-\beta} & \dfrac{1}{\a_{2}-\beta}
\vspace{0.15cm}\\ \dfrac{h(\a_{1})}{\a_{1}-\bar{\beta}} &
\dfrac{h(\a_{2})}{\a_{2}-\bar{\beta}}
\end{array} \right|. \ee

\bigskip

The formulae (1), (3) and (5) permit one to work out a very
concise `canonical' representation of the DN metric. Indeed, using
the results of Ref.~\cite{RMM}, we have \bea &&e=-2(m+\I\nu),
\quad \beta=-m+\I(a-\nu), \nonumber\\ &&f_1=q+\I b, \quad
f(\a_i)=\frac{q+\I b}{\a_i+m-\I(a-\nu)}, \nonumber\\
&&\a_1=-\a_2=\kappa=\sqrt{m^2+\nu^2-a^2-q^2-b^2}. \eea Now,
expanding the determinants in (3), (5) and introducing the
generalized spheroidal coordinates $x$, $y$ via the formulae
\be
2\kappa x=r_2+r_1, \quad 2\kappa y=r_2-r_1, \ee so that $r_1$ and
$r_2$ can be substituted by \be r_1=\kappa(x-y), \quad
r_2=\kappa(x+y), \ee we arrive, after performing some computer
algebra and getting rid of common factors, at the final
expressions for $\E$, $\Phi$, $f$, $\gamma$ and $\omega$: \bea
&&\E=\frac{\kappa x-m-\I(ay+\nu)}{\kappa x+m-\I(ay-\nu)}, \quad
\Phi=\frac{q+\I b}{\kappa x+m-\I(ay-\nu)}, \nonumber\\
&&f=\frac{\kappa^2(x^2-1)-a^2(1-y^2)}{(\kappa x+m)^2+(ay-\nu)^2},
\quad {\rm e}^{2\gamma}=\frac{\kappa^2(x^2-1)-a^2(1-y^2)}
{\kappa^2(x^2-y^2)}, \nonumber\\ &&\omega=2\nu(y-1)-
\frac{a(1-y^2)[2(m\kappa x-\nu ay+m^2+\nu^2)-q^2-b^2]}
{\kappa^2(x^2-1)-a^2(1-y^2)}. \eea

Note that in the new coordinates the line element (4) assumes the
form \bea \d s^2=\kappa^2f^{-1}\left[{\rm e}^{2\gamma}
(x^2-y^2)\left( \frac{\d x^2}{x^2-1}+\frac{\d y^2}{1-y^2}\right)
+(x^2-1)(1-y^2)\d\varphi^2\right] \nonumber\\ -f(\d
t-\omega\d\varphi)^2. \eea

To complete the description of the DN solution in the `canonical'
formalism, it only remains to write out the respective components
of the electromagnetic four--potential. As is well known \cite{E},
the electric component $A_4$ is defined by the real part of the
potential $\Phi$, while the magnetic component $A_3$ is defined by
the real part of Kinnersley's potential $\Phi_2$ \cite{Kin} for
which Sibgatullin's method gives the expression \cite{RMM}
\be
\Phi_2=-\I I/E_-. \ee

From (3), (5) and (10) we then obtain \bea &&A_4=\frac{q(\kappa
x+m)+b(\nu-ay)}{(\kappa x+m)^2+(ay-\nu)^2}, \nonumber\\
&&A_3=b(1-y)+\frac{(1-y)(ay+a-2\nu)[q(\kappa x+m)+b(\nu-ay)]}
{(\kappa x+m)^2+(ay-\nu)^2}. \eea

Therefore, formulae (9), (12) and (7) fully describe the
gravitational and electromagnetic fields in the DN spacetime. It
should be stressed that the parameters $m$, $a$, $\nu$, $q$, $b$
represent an analytically extended parameter set which covers all
the possibilities for a DN massive source to be either a
subextreme object (real $\kappa$), or a superextreme object (pure
imaginary $\kappa$) or an extreme object ($\kappa=0$). This is due
to the fact that the parameters in the axis data (1) represent
five arbitrary and independent multipole moments, as can be easily
seen from (2). Hence, the analysis of the physical properties of
the DN metric does not require in principle the introduction of
the Boyer--Lindquist--like coordinates, and can be carried out
either in the generalized spheroidal coordinates, taking into
account that the product $\kappa x$ is always a real non--negative
quantity, or directly in the Weyl--Papapetrou cylindrical
coordinates $(\rho,z)$. Since, however, the Boyer--Lindquist--like
coordinates $(r,\vartheta)$ which are introduced via the formulae
\be
r=\kappa x+m, \quad \cos\vartheta=y \ee are advantageous for
treating the extreme case, below we write out the DN metric in
these coordinates: \bea \d s^2=\frac{D}{N}\Biggl[ N\Biggl(\frac{\d
r^2}{\Delta}&+&\d\vartheta^2\Biggr)+\Delta\sin^2\vartheta\,\d\varphi^2
\Biggr] \nonumber\\ &-&\frac{N}{D}\Biggl[ \d
t+\Biggl(2\nu(1-\cos\vartheta)
+\frac{aW\sin^2\vartheta}{N}\Biggr)\d\varphi\Biggr]^2, \nonumber
\eea \bea &&D=r^2+(a\cos\vartheta-\nu)^2, \quad
N=r^2-2mr+a^2\cos^2\vartheta-\nu^2+q^2+b^2, \nonumber\\
&&W=2mr+2\nu(\nu-a\cos\vartheta)-q^2-b^2, \nonumber\\
&&\Delta=r^2-2mr-\nu^2+a^2+q^2+b^2. \eea

\section{Some remarks on the limits and physical pro\-perties of
the DN solution}

Since the DN metric is a particular case of the
Pleba\'nski--Demia\'nski solution \cite{PD} the physical and
geometric characteristics of which have been widely discussed in
the literature, in what follows we shall restrict ourselves to
only making some remarks on the properties of the DN metric which
looked to us interesting to be mentioned here.

In the absence of the NUT parameter $\nu$, the DN metric is
asymptotically flat and represents the four--parameter black hole
spacetime involving magnetic charge. This black hole limit,
together with all further possible reductions, was analyzed in
detail by Carter \cite{C}.

The stationary pure vacuum limit ($q=b=0$) is the combined
Kerr--NUT solution. Many authors (see, e.g., \cite{RT,Ar})
associate the DN metric exclusively with the latter
three--parameter vacuum solution, possibly not being aware of the
second part of the paper \cite{DN} where Demia\'nski and Newman
presented their five--parameter electrovacuum metric. A good
discussion of different interpretations of the NUT parameter and
of physical consequences its presence causes in exact solutions
can be found in the review article \cite{Bon}. Here we would only
like to notice that although in a single Kerr--NUT solution the
NUT parameter is non--physical, already in a non--linear
superposition of two Kerr--NUT solutions of Kramer and Neugebauer
\cite{KrN} the formal NUT parameters associated with the two
constituents may give rise to the physical quantities defined via
the multipole moments of the system and, moreover, are necessary
for achieving gravitational equilibrium of the constituents
\cite{TK}. Let us illustrate this with the axis data of the form
\be
e(z)=\frac{z-k-m-\I(a+\nu)}{z-k+m-\I(a-\nu)}\,\cdot
\frac{z+k-m-\I(a-\nu)}{z+k+m-\I(a+\nu)}, \quad f(z)=0 \ee which
can be {\it formally} interpreted, taking into account (1), as
defining a double Kerr--NUT solution whose constituents have equal
masses $m$, angular momenta $a$ and opposite NUT parameters $\nu$
and $-\nu$, the constant $k$ playing a role of the separation
parameter. As a matter of fact, the above axis data represents a
system of identical particles possessing parallel angular momenta.
This can be readily seen from the expressions of the multipole
moments corresponding to (15): \bea &&M_0=2m, \quad M_1=0, \quad
M_2=-2m(m^2+\nu^2-k^2)-2a(ma+2k\nu), \nonumber\\ &&J_0=0, \quad
J_1=2(ma+k\nu), \quad J_2=0, \nonumber\\
&&J_3=-2a^2(ma+3k\nu)-2(3ma+k\nu)(m^2+\nu^2-k^2). \eea

The total NUT parameter of the system is zero ($J_0=0$) and,
therefore, the genuine individual NUT parameters of the
constituents are equal to zero too because of the additional
equatorial symmetry of the system characterized by zero odd
mass--multipoles $M_{2n+1}$ and even angular momentum multipoles
$J_{2n}$, $n=1,2...$ It is important to underline that the formal
NUT parameter $\nu$ in (15) has already nothing to do with either
the physical NUT parameter of the system or with the individual
physical NUT parameters of the constituents; together with the
separation parameter $k$ it defines two physical arbitrary
multipole moments in (16) which are the mass--quadrupole moment
$M_2$ and the angular momentum octupole moment $J_3$.

In general, the DN metric has another non--physical parameter, the
magnetic charge $b$, but like in the case of the Kerr--NUT
solution, this parameter can give rise to physical multipole
moments in the two and many--body systems of aligned, charged,
magnetized, spinning particles. Therefore, the importance of the
DN solution consists in that it provides a single constituent with
the whole set of the parameters which may have physical sense in
the axisymmetric many--body systems of aligned sources.

In view of the above said it would be logic to envisage the
electrovacuum metric \cite{RMM} as describing the non--linear
superposition of $N$ Demia\'nski--Newman solutions.

\section{Conclusion}

The formulae obtained in the present paper give a complete
description of the DN metric and corresponding electromagnetic
potentials within the framework of the Ernst formalism. They can
be used for further investigation of the properties of the DN
spacetime, for example following the line of the recent paper
\cite{Ar}. We have also illustrated that the non--physical
parameters of a single DN solution can acquire physical sense in
the many--body systems.

\section*{Acknowledgements}

We are grateful to Jerzy Pleba\'nski for attracting our attention
to the DN metric in the context of the Ernst formalism, and to
Maciej Przanowski for valuable discussions and for providing us
with important references on the subject. This work was partially
supported by Projects 32138-E and 34222-E from CONACYT of Mexico.

\end{document}